\documentclass[twocolumn,showpacs,preprintnumbers,amsmath,amssymb]{revtex4}

\usepackage{graphicx}
\usepackage{dcolumn}
\usepackage{bm}

\begin{document}

\preprint{Journal of Applied Physics (in press)}

\title{Magnetic-Field and External-Pressure Effect on Ferroelectricity in Manganites: Comparison between GdMnO$_{3}$ and TbMnO$_{3}$}

\author{K. Noda, S. Nakamura, J. Nagayama,}
\author{H. Kuwahara}%
\affiliation{%
Department of Physics, Sophia University\\
Chiyoda-ku, Tokyo 102-8554, Japan}

\date{\today}

\begin{abstract}
We have investigated dielectric properties in Mott insulators GdMnO$_{3}$ and TbMnO$_{3}$ under magnetic fields and external quasihydrostatic pressures. 
In the case of GdMnO$_{3}$, thermal hysteresis for dielectric constant $\varepsilon$ and discontinuous lattice distortion were observed at ferroelectric transition temperature ($T_{C}$), and ferroelectric spontaneous polarization was suppressed by application of external pressure. 
These results indicate that the ferroelectric transition in GdMnO$_{3}$ is a first-order displacive-type one. 
On the other hand, the thermal hysteresis and discontinuous lattice striction were not observed at $T_{C}$ in TbMnO$_{3}$. 
The peak of $\varepsilon$ corresponding to ferroelectric transition was shifted toward higher temperatures by application of external pressure in TbMnO$_{3}$. 
The ferroelectric transition of TbMnO$_{3}$ was thought to be a second-order order-disorder-type one. 
\end{abstract}

\maketitle

\newpage 
\newpage


The strong cross-correlation between nontrivial conjugate variables, such as electric field $E$ vs magnetization $M$ and magnetic field $H$ vs resistivity $\rho$, has been attracted revived interest since the discovery of colossal magnetoresistance (CMR) in manganites.\cite{manganites,Salamon} 
In a localized system in contrast to the itinerant one like CMR manganites, effects of the cross-correlation are observed through a magnetocapacitance or magnetodielectric response in several materials.\cite{katsufuji2,katsufuji1,kimura2} 
Since Kimura $et$ $al$. have recently reported the magnetic control of ferroelectric polarization (so called "magnetic-field-induced electric polarization flop") in TbMnO$_{3}$,\cite{kimura} the system in which (anti)ferromagnetic and ferroelectric properties were strongly connected has been extensively studied and some reports have been published.\cite{hur,goto} 

A series of crystals of Mott insulator $R$MnO$_{3}$ investigated here ($R$ is a trivalent rare earth ion such as La$^{3+}$) is famous as a parent material of the CMR manganites. 
Above-mentioned TbMnO$_{3}$ is one of the series of $R$MnO$_{3}$ which is not a hexagonal but a orthorhombic structure (space group $Pbnm$) with GdFeO$_{3}$-type distortion. 
TbMnO$_{3}$ shows an incommensurate (IC) lattice-modulation wave vector along $b$ axis at sinusoidal antiferromagnetic transition temperature ($T_{N}$$\sim$41K). 
At lower temperatures, ferroelectric phase was observed at the incommensurate-commensurate (or lock-in) magnetic transition ($T_{{\rm lock}}$$\sim$27K). 
It was thought that this ferroelectric transition originates from the sinusoidal lattice modulation induced by the magnetoelastic interaction. 
"Magnetic-field-induced electric polarization flop" observed in TbMnO$_{3}$ was thought to arise from such a strong coupling between magnetism and dielectricity. 
In this case, the direction of ferroelectric polarization is altered from $c$ axis to $a$ by magnetic fields.\cite{kimura} 
As reported in our preceding paper\cite{Kuwahara} the ferroelectric polarization appeared in GdMnO$_{3}$ crystal below 13K, at which temperature Gd 4$f$-spin sublattice antiferromagnetically coupled with respect to Mn 3$d$ spin and the weak ferromagnetism of Mn due to the Dzyaloshinskii-Moriya interaction was suppressed. 
(See also the left panel of Fig.\ref{fig2}.) 
This ferroelectric transition does not correspond to the IC transition ($T_{{\rm IC}}$$\sim$42K) or the $A$-type antiferromagnetic one with weak ferromagnetism ($T_{{\rm N}}$$\sim$20K) in Mn 3$d$ spins. 
The direction of the ferroelectric polarization is along $a$ axis ($P_{a}$) in orthorhombic $Pbnm$ setting, and hystereses reflecting a first-order nature of the transition were observed in thermal and magnetic-field scans. 
These results suggested that this ferroelectric transition was probably due to the lattice modulation connected with the magnetic transition of Gd 4$f$ spins. 
In this work, we clarify the influence of external magnetic fields and quasihydrostatic pressures on the dielectric properties in GdMnO$_{3}$ and TbMnO$_{3}$, and discuss the difference of ferroelectric behavior between them.


Single crystalline samples were grown by the floating zone method. 
We performed x-ray-diffraction measurements on the obtained crystals at room temperature, and confirmed that all crystals show the orthorhombic $Pbnm$ structure without impurity phase. 
All specimens used in this study were cut along crystallographic principal axes into a rectangular shape with a typical dimension of $2.0\times1.5\times0.4$mm$^{3}$\@. 
Measurements of temperature dependence of dielectric constant, spontaneous ferroelectric polarization, and lattice striction in magnetic fields were performed in a temperature-controllable cryostat equipped with a superconducting magnet up to 8T\@. 
Dielectric constant was measured by using LCR meter (Agilent 4284A). 
The spontaneous polarization was obtained by the accumulation of pyroelectric current while the sample was heated at a rate of 4K/min after cooling the sample under a poling field of 500$\sim$300kV/m. 
Lattice striction was measured by using strain gauge and magnetization was measured with a SQUID magnetometer. 
An external quasihydrostatic pressure was produced by a clamp-type piston cylinder cell using Fluorinert as the pressure-transmitting medium.


\begin{figure}
\includegraphics[scale=0.35]{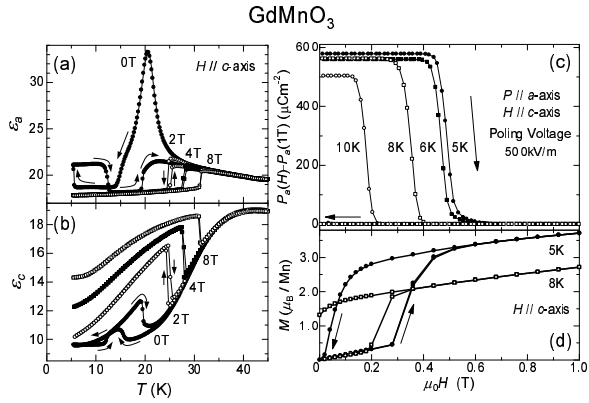}
\vspace{-3mm}
\caption{Temperature dependence of dielectric constant along $a$ axis $\varepsilon_{a}$ (a) and $c$ axis $\varepsilon_{c}$ (b) in magnetic fields (left panel). In Figs. (a) and (b), the thermal hysteresis (concerning the ferroelectric phase) below 13K in 0T completely disappears above 2T while the other hysteresis around 20K in 0T drastically shrinks but still subsists upto 4T. Magnetic field dependence of ferroelectric polarization along $a$ axis $P_{a}$ (c) and magnetization (d) at several fixed temperatures below ferroelectric transition temperature $T_{C}$ (right panel). In these measurements, magnetic field is applied parallel to the $c$ axis. Arrows in the figure represent the scan direction. }
\label{fig1}
\end{figure}

At first, we show in Fig.\ref{fig1} the influence of magnetic fields on the dielectric and magnetic properties in GdMnO$_{3}$. 
As is already shown in our preceding paper,\cite{Kuwahara} ferroelectricity of GdMnO$_{3}$ in zero magnetic field exists below 13K, and is accompanied by the thermal hysteresis in dielectric constant measurement. 
(See also the left panel of Fig.\ref{fig2}.) 
Figures \ref{fig1} (a) and (b) show the temperature dependence of dielectric constant along $a$ axis ($\varepsilon_{a}$) and $c$ ($\varepsilon_{c}$) under magnetic fields parallel to $c$ axis ($H_{c}$) that is perpendicular to the magnetic easy axis ($b$ axis) in GdMnO$_{3}$. The thermal hysteretic region of ferroelectricity was easily suppressed by application of $H_{c}$. 
Figures \ref{fig1} (c) and (d) show the magnetic-field dependence of ferroelectric polarization along $a$ axis ($P_{a}$) and magnetization respectively. 
It turns out that $P_{a}$ was collapsed by application of $H_{c}$ and magnetization was increased simultaneously at the same transition field. 
The observed metamagnetic behavior in $M$-$H$ curves is due to the spin flop of Gd 4$f$ spins coupled to Mn 3$d$ spins antiferromagnetically.\cite{Hamberger} 
These results suggest that ferroelectricity in GdMnO$_{3}$ is strongly correlated to the magnetic order of Gd 4$f$-spins. 
Furthermore, we have also performed similar measurements under magnetic fields parallel to $a$ and $b$ axes. 
The spontaneous polarization $P_{a}$ was suppressed at 2.8T when a magnetic field was applied parallel to $a$ axis ($H_{a}$) while $P_{a}$ did not disappear even if magnetic field was applied up to 8T parallel to the magnetic easy axis of $b$ ($H_{b}$). 
(Not shown in Fig.\ref{fig1}.) 
The ferroelectric transition temperature ($T_{C}$) was shifted toward higher temperature with increase of $H_{b}$, which means that $P_{a}$ was enhanced by application of $H_{b}$. 
These results indicate that the robustness of ferroelectric polarization $P_{a}$ against magnetic fields was sensitive to the magnetic-field direction and $P_{a}$ was strongly coupled with the magnetic structure, in other words, the antiferromagnetic coupling between Gd 4$f$ and Mn 3$d$ spins seems to be crucial to cause $P_{a}$ through lattice modulation. 
Moreover, we have also observed the gigantic magnetocapacitance around $T_{{\rm N}}$ in GdMnO$_{3}$. 
The maximum value of isothermal magnetocapacitance exceeds 68$\%$ at 20K on $\varepsilon_{c}$ under $H_{c}$, where the value of magnetocapacitance is defined as $\Delta$$\varepsilon_{c}$($H$)/$\varepsilon_{c}$($H$$=$$0$) = $[$$\varepsilon_{c}$($H$)$-$$\varepsilon_{c}$($H$$=$0)$]$$/$$\varepsilon_{c}$($H$$=$0).

\begin{figure}
\includegraphics[scale=0.40]{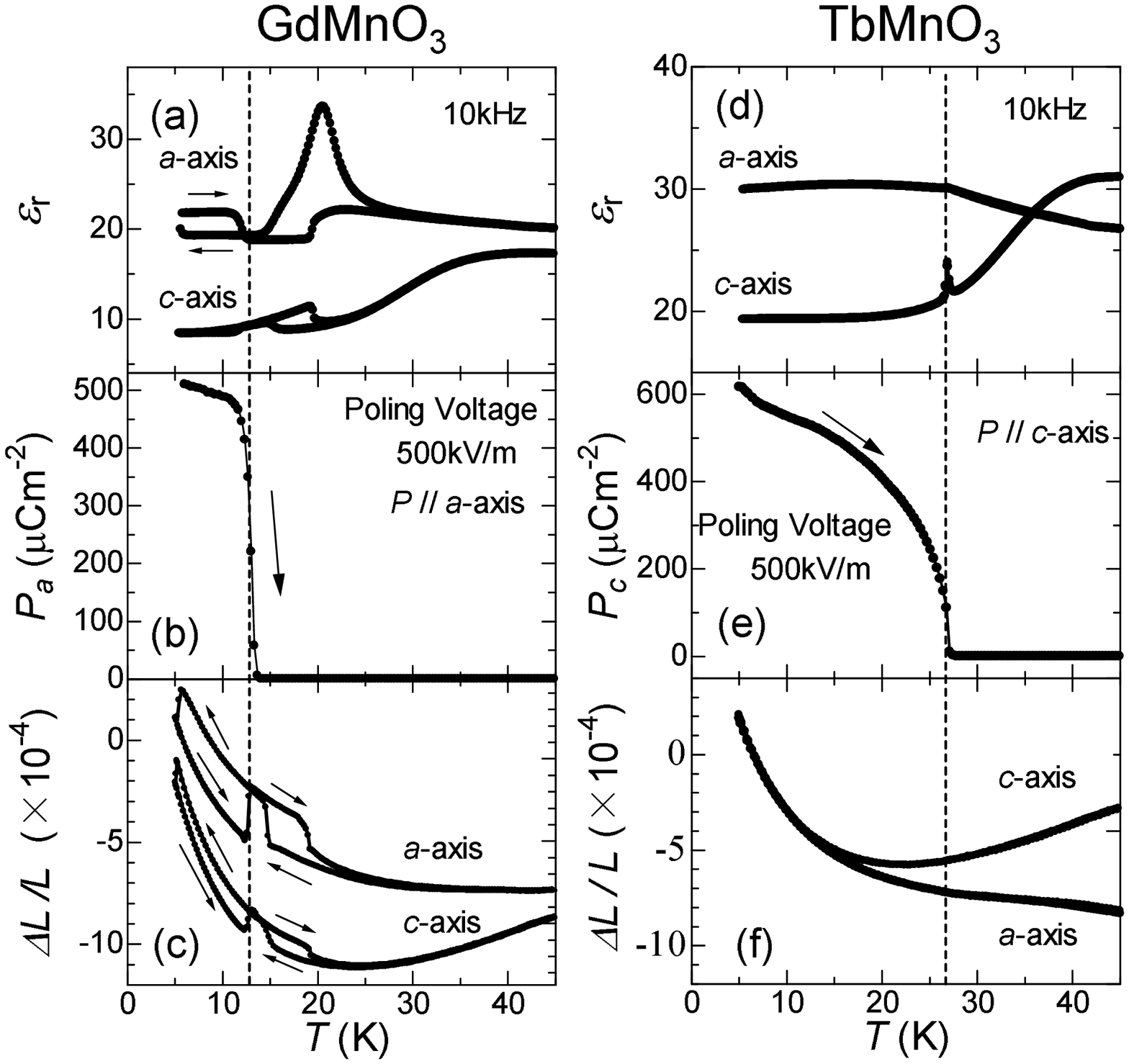}
\vspace{-3mm}
\caption{Temperature dependence of dielectric constant along $a$ and $c$ axes (upper panels), ferroelectric spontaneous polarization (middle), and lattice striction along $a$ and $c$ axes (bottom) in GdMnO$_{3}$ (left panel: a, b, and c) and TbMnO$_{3}$ (right: d, e, and f). Dotted lines indicate the ferroelectric transition temperature. }
\label{fig2}
\end{figure}

Next, we compare the behavior of ferroelectricity in GdMnO$_{3}$ with that of TbMnO$_{3}$, and discuss the difference between them. 
Figure \ref{fig2} shows the temperature dependence of dielectric constant, ferroelectric spontaneous polarization, and lattice striction in GdMnO$_{3}$ (left panel) and TbMnO$_{3}$ (right one) in zero magnetic field. 
As clearly seen in the figure, ferroelectric transition in GdMnO$_{3}$ is accompanied by thermal hystereses and lattice distortion reflecting the nature of the first-order phase transition in contrast to the case of TbMnO$_{3}$. 
In the case of GdMnO$_{3}$, the magnitude of lattice striction coupled with ferroelectric transition was the order of 10$^{-4}$ while no discontinuous jump or drop was observed at $T_{C}$ in TbMnO$_{3}$. 
Moreover, the transition behavior of ferroelectric spontaneous polarization shows the first-order-like abrupt change compared with a second-order-like gradual one in the case of TbMnO$_{3}$. 
The ferroelectric transition of GdMnO$_{3}$ is the first-order one, on the other hand, it is thought that the ferroelectric transition of TbMnO$_{3}$ is the second-order one. 
We carried out curve fitting of spontaneous polarization $P_{c}$ in TbMnO$_{3}$ defined as $P_{c}$$(T)$/$P_{c}$($T$$=$$0$) $=$ $[$($T_{C}$$-$$T$)/$T_{C}$$]$$^{\beta}$, because of its second-order like transition. 
The obtained value of the critical exponent $\beta$ was 0.365 which agreed well with an expected value for the second-order ferromagnetic transition in the three dimensional Heisenberg model.\cite{Holm} 
This result also supports that the ferroelectric transition of TbMnO$_{3}$ is the second-order one, while the first-order transition in GdMnO$_{3}$. 
This conclusion is supported by other measurements including external pressure experiment discussed in the following paragraph.

\begin{figure}
\includegraphics[scale=0.34]{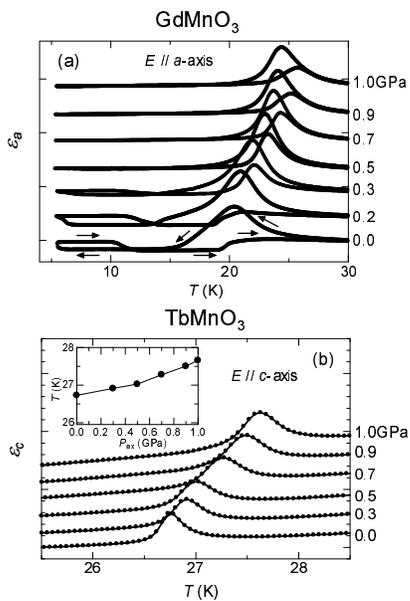}
\vspace{-3mm}
\caption{Temperature dependence of dielectric constant under several fixed external quasihydrostastic pressures $P_{{\rm ex}}$ in GdMnO$_{3}$ (a) and TbMnO$_{3}$ (b). Note that dielectric constants in GdMnO$_{3}$ and TbMnO$_{3}$ are measured along $a$ and $c$ axes respectively. Vertical axes in (a) and (b) are shifted for clarity. Inset in (b) shows $P_{{\rm ex}}$ dependence of ferroelectric transition temperature $T_{C}$ of TbMnO$_{3}$. }
\label{fig3}
\end{figure}

Furthermore, in order to clarify the difference in mechanism of ferroelectricity between GdMnO$_{3}$ and TbMnO$_{3}$, we have performed measurements of dielectric constant under external quasihydrostatic pressure $P_{{\rm ex}}$. 
Figure \ref{fig3} shows the temperature dependence of dielectric constant of GdMnO$_{3}$ (a) and TbMnO$_{3}$ (b) under several fixed pressures. 
In the case of GdMnO$_{3}$, the lower thermal hysteretic region below 13K corresponding to the ferroelectric phase was suppressed with increasing pressure and it disappears above 0.5 GPa. 
It seems that $T_{C}$ was shifted toward lower temperatures with increasing pressure and it reached the lower temperatures that we can not obtain. 
The tendency that $T_{C}$ shifts toward lower temperatures with increasing pressure is a typical characteristic of a displacive-type ferroelectric transition under external pressure.\cite{Samara} 
The obtained tendency, therefore, implies that the ferroelectric transition of GdMnO$_{3}$ is the displacive-type one. 
In the perovskite structure, many displacive-type-ferroelectric materials such as BaTiO$_{3}$, PbTiO$_{3}$, LiTaO$_{3}$, and so on have actually been reported. 
Then it seems reasonable to conclude that the ferroelectric transition of GdMnO$_{3}$ is the first-order displacive-type one. 
On the other hand, in the case of TbMnO$_{3}$, the dielectric constant peak corresponding to the ferroelectric transition is shifted toward higher temperatures with increasing pressure. 
This behavior is a typical characteristic of a order-disorder-type ferroelectric transition under external pressures.\cite{Samara}  
This result also supports that the ferroelectric transition of TbMnO$_{3}$ is the second-order order-disorder-type one. 

Although GdMnO$_{3}$ and TbMnO$_{3}$ are nearly same crystal structure with slight different orthorhombic distortion,\cite{kimura2} there may exist similar or distinct behavior of ferroelectricity between them. 
In TbMnO$_{3}$, the ferroelectric polarization along $a$ axis ($P_{a}^{{\rm Tb}}$($H$$\neq$0)) is flopped from $c$ axis ($P_{c}^{{\rm Tb}}$($H$$=$0) induced by magnetic fields. 
$P_{a}^{{\rm Tb}}$($H$$\neq$0) accompanied by hysteresis was observed, only when the magnetic field was applied parallel to $b$ axis. 
The spontaneous polarization of GdMnO$_{3}$ along $a$ axis in zero field ($P_{a}^{{\rm Gd}}$($H$$=$0)) was enhanced only when the magnetic field was applied parallel to the same $b$ axis, and it was accompanied by hysteresis at $T_{C}$. 
As discussed in previous paragraph, transition character between GdMnO$_{3}$ and TbMnO$_{3}$ in zero magnetic field is quite different from each other. 
However, as mentioned above, there are many common features between $P_{a}^{{\rm Tb}}$($H$$\neq$0) and $P_{a}^{{\rm Gd}}$($H$=0). 
These facts may imply that $P_{a}^{{\rm Tb}}$($H$$\neq$0) has a similar origin of $P_{a}^{{\rm Gd}}$($H$$=$0), which cannot be revealed in detail at present, although $P_{c}^{{\rm Tb}}$($H$$=$0) seems to have quite different origin from $P_{a}^{{\rm Gd}}$($H$$=$0).

In summary, we have studied the influence of magnetic field and quasihydrostatic pressure upon the dielectric properties in single crystals of GdMnO$_{3}$ and TbMnO$_{3}$. 
Contrast to the case of TbMnO$_{3}$, "magnetic-field-induced electric polarization flop" was not observed in GdMnO$_{3}$. 
The robustness of polarization in GdMnO$_{3}$ $P_{a}^{{\rm Gd}}$ against magnetic fields has the large anisotropy respect to the field direction. 
This indicates that $P_{a}^{{\rm Gd}}$ was strongly reflected by the large anisotropy of magnetic structure, i.e., antiferromagnetic coupling between Gd 4$f$ and Mn 3$d$ spins. 
The collapse of $P_{a}^{{\rm Gd}}$ seems to arise from change of lattice modulation coupled magnetoelastically with the metamagnetic transition of Gd 4$f$ spins. 
By comparing the experimental results in the cases of GdMnO$_{3}$ and TbMnO$_{3}$, the transition character of $P_{a}^{{\rm Gd}}$($H$$=$0) seems to be similar to $P_{a}^{{\rm Tb}}$($H$$\neq$0) which is flopped from $P_{c}^{{\rm Tb}}$($H$$=$0) by application of magnetic field parallel to the $b$ axis $H_{b}$. 
On the other hand, the origin of $P_{a}^{{\rm Gd}}$($H$$=$0) and $P_{c}^{{\rm Tb}}$($H$$=$0) is likely to be different: The former should be classified as the first-order displacive-type transition and the latter as the second-order order-disorder-type one. 
The observed coupling between ferroelectric and magnetic properties is expected to bring a possible multifunctional device with cross-correlation such as magnetically-recorded ferroelectric memory in the future.


\newpage

\newcounter{volume}
\setcounter{volume}{99}

\newpage

\end{document}